\documentclass[conference]{IEEEtran}
\IEEEoverridecommandlockouts
\usepackage{cite}
\usepackage{amsmath,amssymb,amsfonts}
\usepackage{algorithmic}
\usepackage{graphicx}
\usepackage{textcomp}
\usepackage{xcolor}
\usepackage{stfloats}
\def\BibTeX{{\rm B\kern-.05em{\sc i\kern-.025em b}\kern-.08em
    T\kern-.1667em\lower.7ex\hbox{E}\kern-.125emX}}
\begin{document}

\title{Information-Theoretic Analysis of Brain MRI: Mutual Information and Pixel Intensity Patterns in Tumor vs. Normal Tissues\\}

\author{\IEEEauthorblockN{1\textsuperscript{st} Mazaher Kabiri}
\IEEEauthorblockA{\textit{Dept. of industrial engineering } \\
\textit{Binghamton University}\\
Binghamton, Ny, USA \\
mkabiri@binghamton.edu}
\and
\IEEEauthorblockN{2\textsuperscript{nd}  Shahd Qasem Mah'd Tarman}
\IEEEauthorblockA{\textit{Dept. of industrial engineering } \\
\textit{Binghamton University }\\
Binghamton, NY, USA \\
starman@binghamton.edu}
}

\maketitle

\begin{abstract}
The application of information theory in medical imaging, particularly in magnetic resonance imaging (MRI), offers powerful quantitative tools for analyzing structural differences in brain tissues. This study utilizes mutual information (MI) and pixel intensity distributions to differentiate between normal and tumor-affected brain MRI images. Mutual information analyses revealed significantly higher MI values in tumor images compared to normal ones, indicating greater internal similarity within tumor images. Pixel intensity analysis further demonstrated distinct distribution patterns between the two groups: tumor images showed pronounced pixel frequency concentrations within a specific intensity range (0.3–0.4), suggesting predictable structural characteristics. Conversely, normal images exhibited broader, more uniform pixel intensity distributions across most intensity ranges, except for an initial peak observed in both groups. These findings highlight the capability of information-theoretic metrics, such as mutual information and pixel intensity analysis, to effectively distinguish tumor tissue from normal brain structures, providing promising avenues for enhanced diagnostic and analytical methods in neuroimaging.
\end{abstract}

\renewcommand\IEEEkeywordsname{Keywords}

\section{Introduction}
Magnetic Resonance Imaging (MRI) has become a cornerstone in the diagnosis and monitoring of neurological conditions, particularly brain tumors\cite{bauer2013survey}. Traditional radiological assessments rely heavily on qualitative interpretations of structural abnormalities, which may introduce subjectivity\cite{erickson2017machine} and overlook subtle yet informative patterns. In recent years, computational approaches rooted in information theory have emerged as powerful tools to quantify image complexity and extract latent features\cite{cover2006elements} that may not be visually apparent.\\
Information theory, originally developed to study the transmission of data, provides a mathematical framework for measuring uncertainty and information content in signals. When applied to medical imaging, metrics such as mutual information (MI) and entropy-based pixel intensity distributions\cite{pluim2003mutual} can reveal underlying statistical structures and similarities between images or regions of interest. These measures are particularly relevant in brain imaging, where tumor tissues often present irregular, yet internally consistent, patterns\cite{tustison2014optimal} that differ markedly from normal brain anatomy. Notably, brain tissue’s movement, dilation, and contraction significantly influence cerebrospinal fluid (CSF) dynamics and may subtly impact image-based measurements\cite{shojaeianforoud2025fluid}, underscoring the importance of incorporating physiological context in computational imaging analyses.\\
This study investigates the utility of mutual information and pixel intensity histograms in differentiating tumor-affected brain MRI images from normal ones. By comparing the internal consistency of tumor images to the variability observed in healthy brain tissue, we aim to demonstrate how information-theoretic analysis can enhance diagnostic interpretation. Our results indicate that tumor images exhibit higher mutual information values\cite{lopez2012mutual} and a more concentrated pixel intensity range, suggesting the presence of predictable structural redundancies within tumor regions. These findings underscore the potential of information theory as a complementary quantitative approach in neuroimaging and tumor assessment. 

\section{Background}
Information theory, originally developed by Claude Shannon\cite{shannon1948}, in the context of communication systems, provides mathematical tools to quantify uncertainty, redundancy, and information content\cite{cover2006elements} in signals. In the realm of medical imaging, these principles have been adapted\cite{pluim2003mutual} to analyze structural and statistical properties of images, offering an objective and quantitative complement to conventional visual assessments.\\
Two core concepts in information theory—mutual information (MI) and entropy-based pixel intensity analysis—are particularly relevant to brain MRI data. Entropy measures the average unpredictability or complexity of a signal\cite{tsai1995fast}. In image analysis, it captures how pixel intensities are distributed across an image. A more uniform or widespread distribution generally indicates higher entropy, suggesting greater randomness or structural variability. Conversely, low entropy may reflect concentrated or predictable intensity patterns.\\
Mutual Information (MI) quantifies the amount of information\cite{maes1997multimodality} that one image shares with another. In the context of medical imaging, it can measure the similarity or dependency between regions within or across different images. High MI values between tumor-affected brain images indicate that these images share significant internal structural similarities—potentially due to repetitive morphological features of tumor tissue. Lower MI values in normal brain images may suggest more heterogeneous structural organization, which is typical in healthy anatomical variability.\\
Pixel intensity histograms offer another lens into image structure\cite{gonzalez2008digital}. By plotting the frequency of each pixel intensity level, one can observe whether certain intensity ranges are more dominant or evenly distributed. Tumor images often show intensity peaks\cite{bauer2013survey} in specific narrow bands, implying localized homogeneity, while normal images usually display a more balanced distribution across a wider intensity range.\\
This theoretical framework underpins the analyses performed in this study, where brain MRI scans were examined using mutual information and pixel intensity histograms to detect and characterize differences between tumor and non-tumor tissue. These metrics allow for a data-driven, quantifiable approach to structural interpretation in neuroimaging.
\section{Materials and Methods}
The dataset used in this study was obtained from Kaggle, a publicly accessible data repository. It contains a total of 40 brain MRI images, divided equally into two categories: 20 images from patients diagnosed with brain tumors (tumor group) and 20 images representing healthy individuals (normal group). All images were provided in standard image formats (e.g., JPG or PNG) and were grayscale or converted to grayscale prior to analysis.\\
To ensure consistency during processing, each image was resized to a uniform resolution of 256×256 pixels. The dataset was manually reviewed to confirm proper labeling and visual clarity. Figure 1 displays a representative example from each class.\\
\begin{figure}[htbp]
  \centering
  \includegraphics[width=\linewidth]{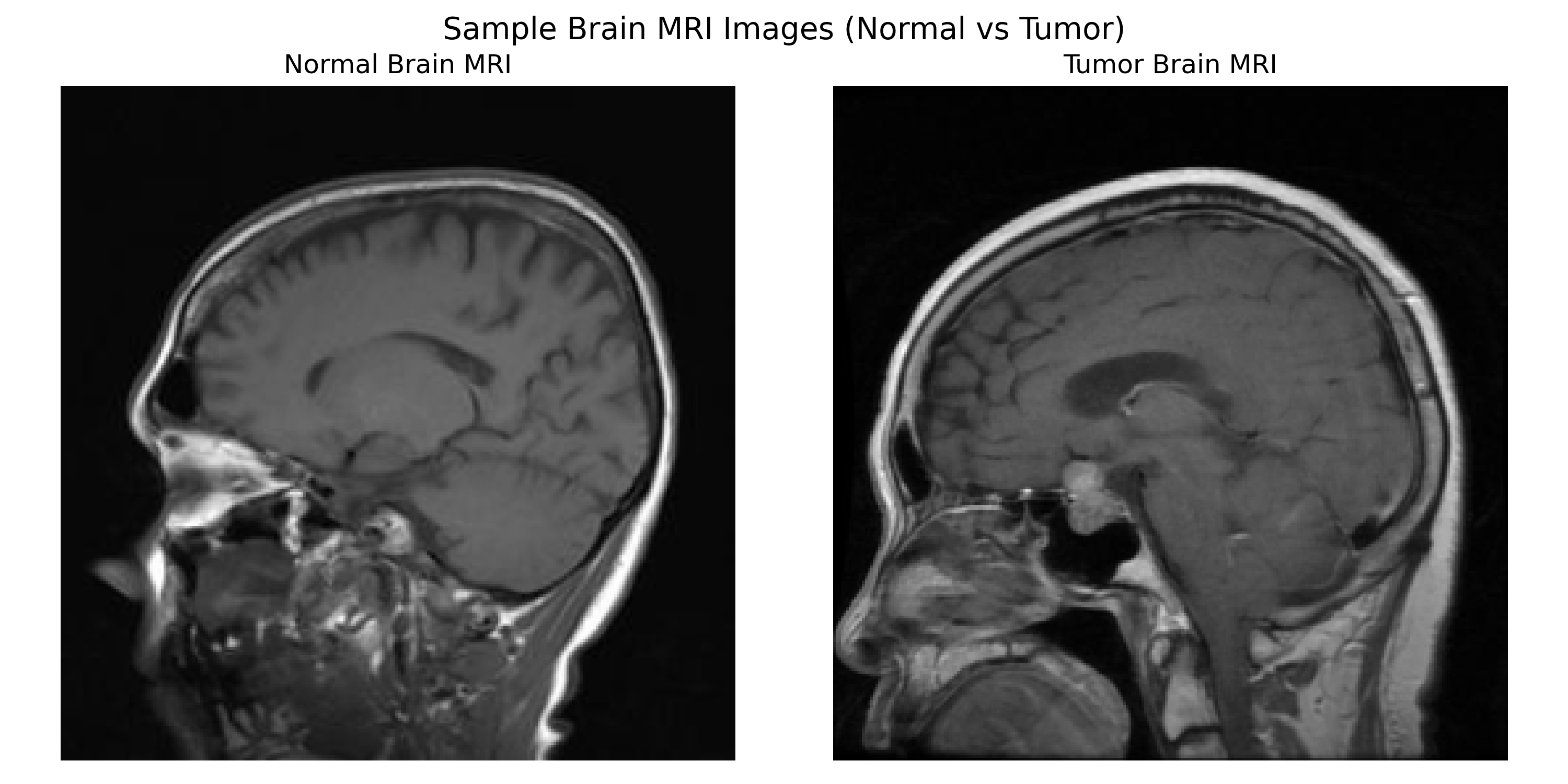}
  \caption{Sample brain MRI images. Left: Normal brain; Right: Tumor-affected brain.}
\end{figure}

\section*{Information entropy}

Self-information quantifies the amount of information contained in the realization of a specific value of a random variable. Given a discrete random variable \( X \) with probability distribution \( P(X) \), where \( X \) takes values in the set \( \mathcal{X} = \{x_1, x_2, \dots, x_n\} \), the self-information is defined as:

\begin{equation}
I(X) = - \sum_{x \in \mathcal{X}} P(x) \log P(x)
\end{equation}

This expression represents the entropy of the random variable \( X \), indicating the average level of uncertainty inherent in the distribution.

\vspace{1em}

\section*{Mutual Information}

Mutual information measures the amount of information obtained about one random variable through the observation of another. Given two random variables \( X \) and \( Y \) with a joint distribution \( P(X, Y) \), mutual information is defined as:

\begin{equation}
I(X; Y) = \sum_{x \in \mathcal{X}} \sum_{y \in \mathcal{Y}} P(x, y) \log \frac{P(x, y)}{P(x) P(y)}
\end{equation}

Alternatively, it can also be written as:

\begin{equation}
I(X; Y) = \sum_{x \in \mathcal{X}} \sum_{y \in \mathcal{Y}} P(x, y) \log \frac{P(x \mid y)}{P(x)}
\end{equation}

This leads to the identity:

\begin{equation}
I(X; Y) = I(X) - I(X \mid Y)
\end{equation}

Where:
- \( I(X \mid Y) \) denotes the conditional entropy of \( X \) given \( Y \).
- \( P(x \mid y) \) is the conditional probability of \( X \) given \( Y \).

Mutual information is always non-negative and equals zero if and only if \( X \) and \( Y \) are statistically independent.

\section*{Pixel Intensity and Histogram Analysis}

Pixel intensity refers to the brightness value assigned to each pixel in a grayscale image. In digital medical imaging, especially MRI scans, these intensity values typically range from 0 (black) to 255 (white) in 8-bit grayscale representation. The distribution of pixel intensities across an image reflects the structural and textural characteristics of the tissue being imaged.

To analyze this distribution, a \textit{pixel intensity histogram} is constructed. This histogram is a graphical representation where the x-axis corresponds to intensity levels and the y-axis represents the number of pixels (frequency) with each specific intensity.

The histogram $H(i)$ for an image of size $M \times N$ is computed by counting the number of times each intensity value $i \in [0, 255]$ appears in the image:

\begin{equation}
H(i) = \sum_{x=1}^{M} \sum_{y=1}^{N} \delta(I(x, y) = i)
\tag{5}
\end{equation}

Where:
\begin{itemize}
  \item $I(x, y)$ is the intensity value at pixel location $(x, y)$.
  \item $\delta$ is an indicator function: it returns 1 if the condition is true, and 0 otherwise.
\end{itemize}

The histogram can be normalized to represent a probability distribution by dividing each bin count by the total number of pixels:

\[
P(i) = \frac{H(i)}{M \times N} \tag{6}
\]

This normalized histogram $P(i)$ is especially useful for statistical analysis, entropy calculation, and comparison between different images regardless of their size.

\section*{Libraries}
The analysis was implemented in Python using several standard scientific libraries. OpenCV was employed for basic image processing tasks such as image loading, grayscale conversion, and resizing. NumPy was used for array manipulation, histogram computation, and normalization of pixel intensity distributions. Matplotlib facilitated the visualization of images and histograms. For information-theoretic analysis, including entropy computation, core NumPy functions were used. Scikit-learn was utilized to calculate mutual information between image pairs using its mutual information score function.

\section{Results}
\section*{Mutual information}
\begin{figure*}[!htbp]
    \centering
    \includegraphics[width=\textwidth]{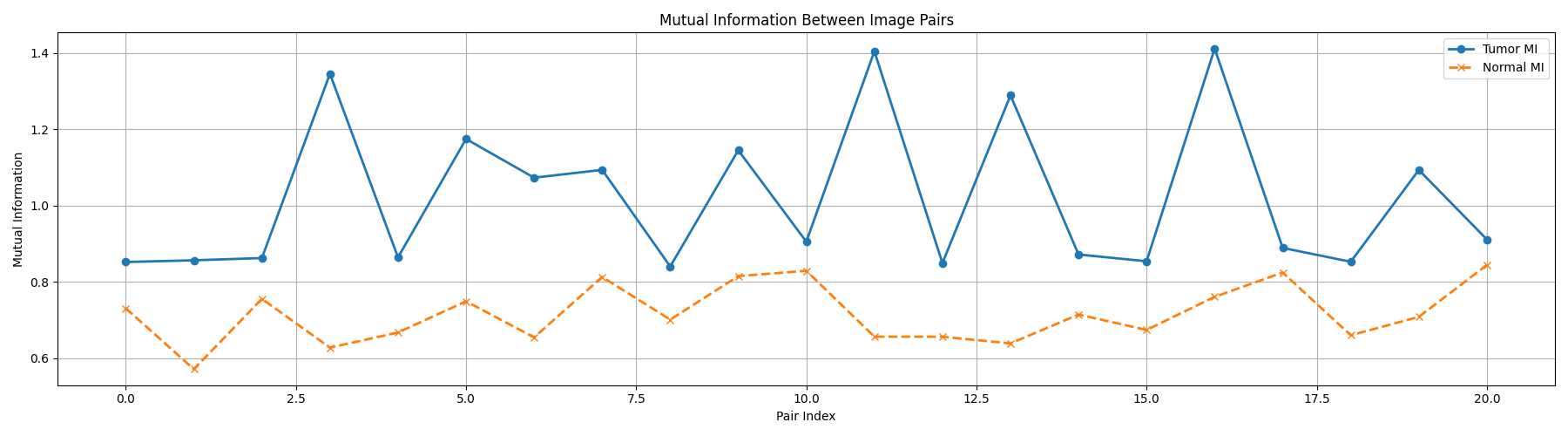}
    \caption{Mutual information values between MRI image pairs.}
\end{figure*}

Figure 2 presents the mutual information (MI) values computed between pairs of tumor and normal brain MRI images. The solid blue line represents MI values for tumor image pairs, while the dashed orange line corresponds to normal image pairs. Each point along the x-axis refers to a unique image pair index, and the y-axis shows the mutual information value for that pair.

Overall, tumor image pairs exhibit higher MI values than normal pairs across most indices. Tumor MI values frequently exceed 1.0 and reach up to approximately 1.4, indicating a high degree of shared structural information between tumor-affected images. These peaks occur repeatedly, reflecting consistent similarity patterns in tumor morphology.

In contrast, normal image pairs maintain relatively stable MI values, generally ranging between 0.6 and 0.85. The flatter trajectory of the orange dashed line implies lower and more uniform structural similarity among normal brain images, which is consistent with expected anatomical variability in healthy brains.

The contrast in MI behavior between tumor and normal groups highlights the distinct internal redundancy present in tumor images. The repeated occurrence of high MI spikes in the tumor group suggests that tumors exhibit characteristic intensity patterns that recur across samples, while normal images remain more heterogeneous.
\section*{Pixel intensity}
\begin{figure*}[t]
    \centering
    \includegraphics[width=\textwidth]{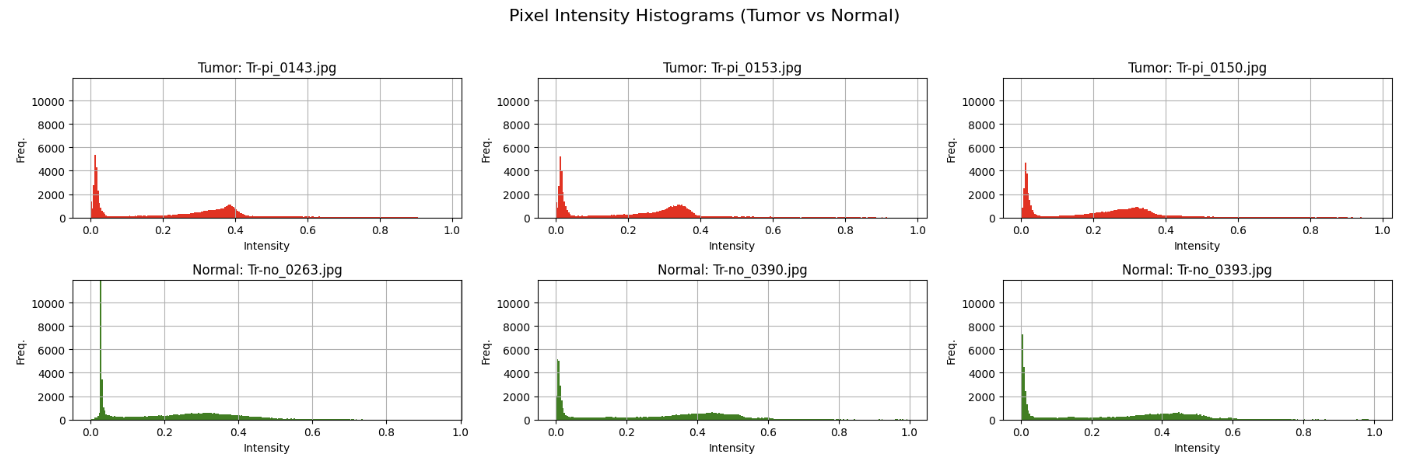}
    \caption{Pixel intensity histogram 256*256.}
\end{figure*}

Figure 3 displays the pixel intensity histograms of selected tumor and normal brain MRI images. The top row shows histograms for three tumor samples, while the bottom row presents their corresponding normal counterparts. The x-axis represents normalized pixel intensity values (ranging from 0 to 1), and the y-axis indicates the frequency of pixels at each intensity level.

Tumor images exhibit pixel intensity distributions that are more concentrated within specific mid-low intensity ranges, particularly around 0.1 to 0.4. While there is an initial peak near zero in all samples, tumor histograms consistently show additional peaks and elevated frequencies in a secondary band (~0.3–0.4), suggesting repeated intensity patterns associated with pathological structures. This secondary peak is more distinct and consistently visible across all tumor cases shown.

In contrast, normal image histograms demonstrate a sharper, singular spike near zero intensity with minimal secondary structure. The distributions for normal scans are more uniform after the initial spike, lacking the mid-range clustering observed in tumor images. This reflects the broader anatomical variability and the absence of localized abnormal intensity clusters in healthy tissue.

The observed patterns suggest that tumor tissues tend to exhibit more predictable and repeated intensity characteristics, possibly due to consistent lesion composition or imaging properties across cases. Normal tissues, on the other hand, display broader dispersion of intensities and lack the secondary peaks seen in tumor cases, reinforcing their heterogeneity.

\section{Discussion}
In this study, pixel intensity distributions were initially analyzed on MRI images resized to a fixed resolution of 256×256 pixels\cite{bauer2013survey}. To evaluate the robustness of intensity-based patterns across different image scales, additional experiments were conducted using alternative resolutions:\cite{liao2015multi} 64×64, 128×128, and 512×512 pixels. Figures 4 to 6, respectively. The resulting histograms for each resolution showed consistent structural trends\cite{kumar2012radiomics}, particularly the concentration of tumor image intensities in the mid-low range and the broader dispersion observed in normal images.

This resolution-invariant behavior supports the reliability of pixel intensity analysis\cite{tustison2014optimal} as a low-level yet informative feature for distinguishing between tumor and normal brain structures. While the shape and smoothness of the histograms varied slightly due to downsampling or upsampling effects, the general patterns remained recognizable across all resolutions.
Despite these findings, the current study has several limitations. First, the dataset used was relatively small\cite{esteva2017dermatologist}, comprising only 40 images equally split between tumor and normal cases. This limited sample size may restrict the generalizability of the results to broader clinical datasets. Second, the analysis focused solely on grayscale intensity information and mutual information\cite{haralick1973textural} metrics, without incorporating spatial or texture-based features, which could provide additional discriminative power. Additionally, ground-truth segmentations of tumor regions were not available, so the analysis was applied to entire brain slices rather than localized regions of interest.

Future work could expand on this analysis by including larger and more diverse MRI datasets, ideally with voxel-level annotations for precise region-based evaluation. Integrating texture descriptors, frequency-domain features\cite{shen2017deep} (e.g., wavelet transforms), or deep learning-based feature extraction methods may enhance classification performance. Furthermore, a multi-resolution framework that combines pixel-level intensity statistics with higher-level semantic features could lead to a more comprehensive understanding of tumor vs. normal brain structure in medical imaging.
\begin{figure}[!t]  % برای یک ستونه
    \centering
    \includegraphics[width=1\linewidth]{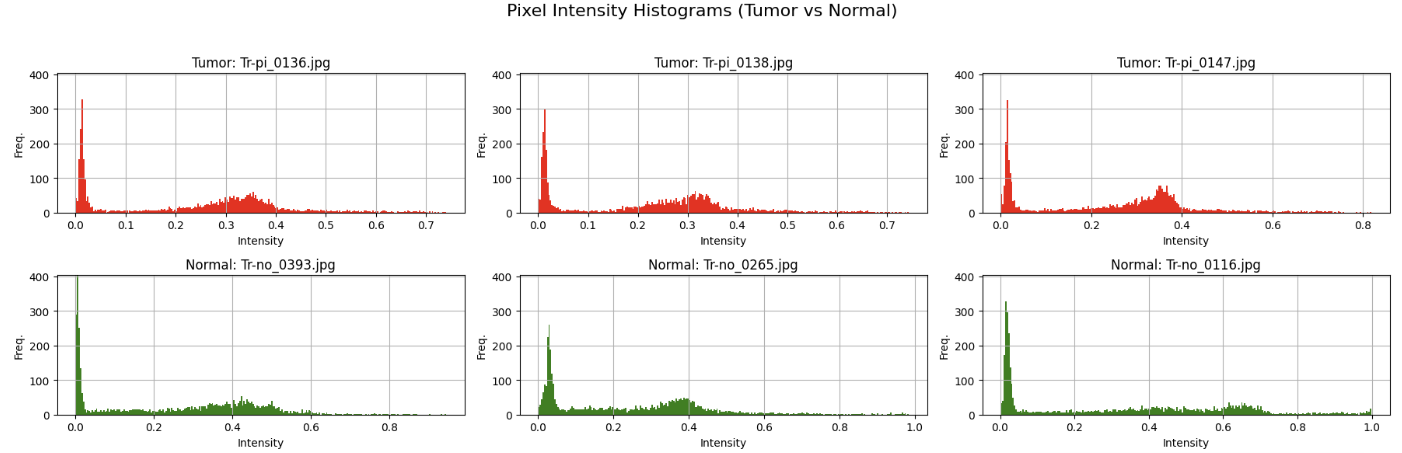}
    \caption{Pixel intensity histogram 64*64.}
\end{figure}
\begin{figure}[!t]  % برای یک ستونه
    \centering
    \includegraphics[width=1\linewidth]{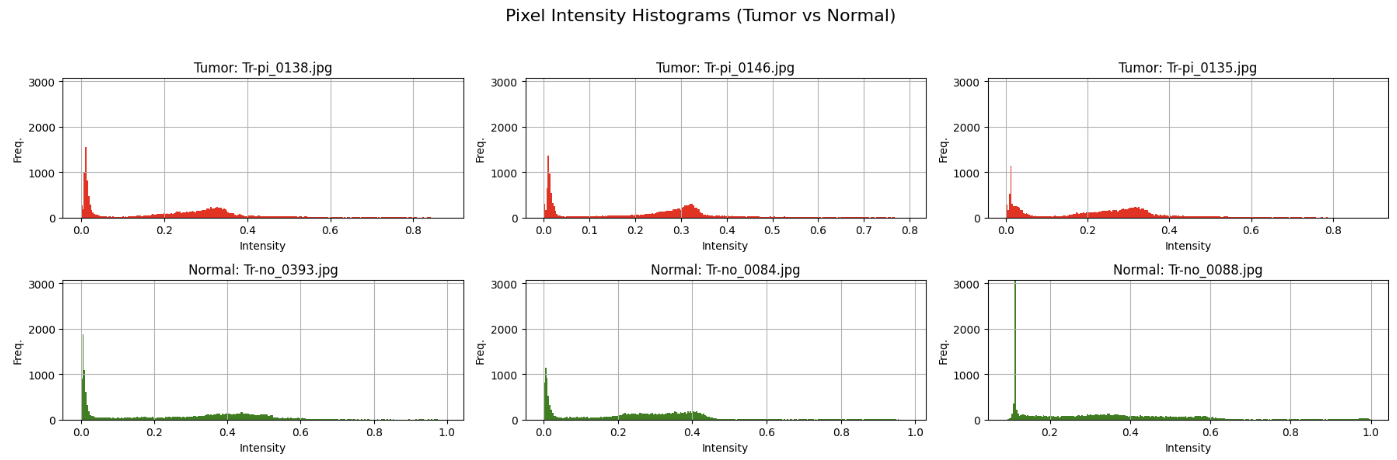}
    \caption{Pixel intensity histogram 128*128.}
\end{figure}
\begin{figure}[!t]  % برای یک ستونه
    \centering
    \includegraphics[width=1\linewidth]{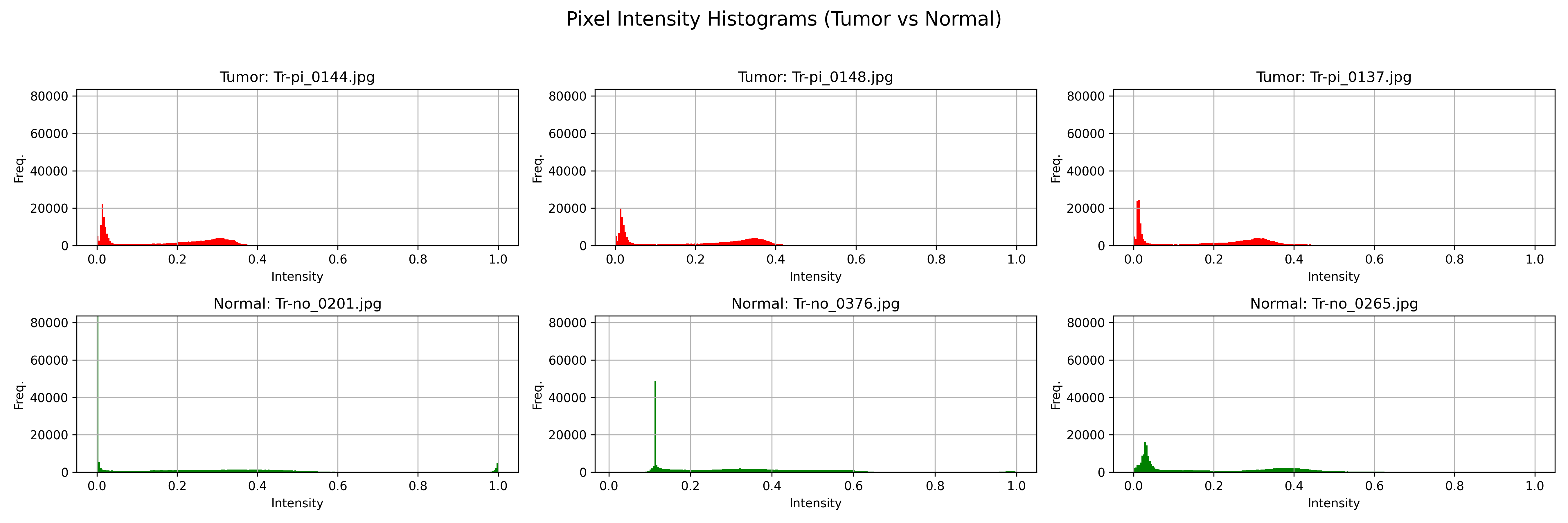}
    \caption{Pixel intensity histogram 512*512.}
\end{figure}
\section{Conclusion}
This study demonstrates that both mutual information and pixel intensity distribution are effective quantitative features for distinguishing between tumor and normal brain MRI images. Tumor images consistently exhibited higher mutual information values, indicating greater internal structural redundancy, while normal images showed lower and more uniform MI, reflecting natural anatomical variability.

Pixel intensity histograms further supported this distinction: tumor images displayed concentrated intensity clusters—especially within the mid-low intensity range—while normal images exhibited a broader and flatter distribution. These patterns were preserved across various image resolutions (64×64, 128×128, 256×256, and 512×512), suggesting that both mutual information and pixel intensity analysis are robust to scale and can serve as reliable low-level indicators of abnormal brain structure.

Taken together, the results suggest that information-theoretic and intensity-based features offer a promising foundation for simple yet effective brain tumor detection frameworks, particularly in data-limited or early-stage diagnostic scenarios.
\section{Authors’ Contributions}
The author solely conducted all aspects of the study. This includes conceptualizing the research idea, collecting and preprocessing the MRI data, performing mutual information and pixel intensity analyses, interpreting the results, generating visualizations, and drafting the manuscript. All experiments, coding, and documentation were carried out independently.
\section{References}

\bibliographystyle{IEEEtran}
\bibliography{ref}
\section*{List of Equations with References}
\begin{itemize}
    \item Equation (1): Entropy of an Image — Shannon (1948)~\cite{shannon1948}
    
    \item Equation (2): Mutual Information (basic definition) 
    \textit{Pluim et al. (2003)~\cite{pluim2003mutual}}

    \item Equation (3): MI via Conditional Probability   
    \textit{Cover \& Thomas (2006)~\cite{cover2006elements}}
    \item Equation (4): MI via Entropy Identity   
    \textit{Cover \& Thomas (2006)~\cite{cover2006elements}}
    \item Equation (5): Pixel Intensity (grayscale image)  
    \textit{Gonzalez \& Woods (2008)~\cite{gonzalez2008digital}}
    \item Equation (6): Histogram Normalization   
    \textit{Gonzalez \& Woods (2008)~\cite{gonzalez2008digital}}
\end{itemize}

\end{document}